\documentclass[useAMS,usenatbib,letterpaper]{mn2e}
\usepackage{graphicx}
\usepackage{amsmath}
\usepackage{amssymb}

\newcommand{\simgt}{\hbox{\rlap{\raise 0.425ex\hbox{$>$}}\lower
    0.65ex\hbox{$\sim$}}} \newcommand{\simlt}{\hbox{\rlap{\raise
      0.425ex\hbox{$<$}}\lower 0.65ex\hbox{$\sim$}}}

\title[Mass transport by buoyant bubbles in galaxy
  clusters]{Mass transport by buoyant bubbles in galaxy
  clusters}

\author[E.C.D. Pope, A. Babul, G.Pavlovski, R.G. Bower, A. Dotter]
       {Edward C.D. Pope$^{1}$\thanks{E-mail:ecdpope@uvic.ca}, Arif
         Babul$^{1}$, Georgi Pavlovski$^{2}$, Richard G. Bower$^{3}$,
         Aaron Dotter$^{1}$\\$^{1}$Department of Physics \& Astronomy,
         University of Victoria, Victoria, BC, V8P 1A1,
         Canada\\$^{2}$School of Physics \& Astronomy, University of
         Southampton, Southampton, SO17 1BJ, UK\\$^{3}$Institute for
         Computational Cosmology, Department of Physics, Durham
         University, South Road, Durham DH1 3LE, UK\\}

\begin{document}

\pagerange{\pageref{firstpage}--\pageref{lastpage} \pubyear{2010}}

\maketitle

\label{firstpage}

\begin{abstract}
We investigate the effect of three important processes by which
AGN-blown bubbles transport material: drift, wake transport and
entrainment. The first of these, drift, occurs because a buoyant
bubble pushes aside the adjacent material, giving rise to a net upward
displacement of the fluid behind the bubble. For a spherical bubble,
the mass of upwardly displaced material is roughly equal to half the
mass displaced by the bubble, and should be $\sim 10^{7-9}\,{\rm
  M_\odot}$ depending on the local ICM and bubble parameters. We show
that in classical cool core clusters, the upward displacement by drift
may be a key process in explaining the presence of filaments behind
bubbles. A bubble also carries a parcel of material in a region at its
rear, known as the wake. The mass of the wake is comparable to the
drift mass and increases the average density of the bubble, trapping
it closer to the cluster centre and reducing the amount of heating it
can do during its ascent. Moreover, material dropping out of the wake
will also contribute to the trailing filaments. Mass transport by the
bubble wake can effectively prevent the build-up of cool material in
the central galaxy, even if AGN heating does not balance ICM cooling.
Finally, we consider entrainment, the process by which ambient
material is incorporated into the bubble. Studies of observed bubbles
show that they subtend an opening angle much larger than predicted by
simple adiabatic expansion. We show that bubbles that entrain ambient
material as they rise, {\em will} expand faster than the adiabatic
prediction; however, the entrainment rate required to explain the
observed opening angle is large enough that the density contrast
between the bubble and its surroundings would disappear rapidly. We
therefore conclude that entrainment is unlikely to be a dominant mass
transport process. Additionally, this also suggests that the bubble
surface is much more stable against instabilities that promote
entrainment than expect for pure hydrodynamic bubbles.
\end{abstract}

\begin{keywords}

\end{keywords}

\section{Introduction}
The hot, gaseous atmospheres of galaxy clusters often show depressions
in the X-ray surface brightness \citep[see, for
  example][]{mc02,birzan,dunn05,rafferty06}. These depressions are
indicative of empty cavities, or bubbles, embedded in the hot gas
\citep[e.g.][]{mcnuls}. The presence of bubbles is generally taken to
be a signature of AGN feedback. In this model, a fraction of the
material cooling from the gaseous atmosphere is accreted by a
supermassive black hole located in the central galaxy. This releases
vast amounts of energy often in the form of outflows which couple to
the hot atmosphere \citep[e.g.][]{calori,babul}.

AGN feedback is widely considered to have important consequences for
the evolution of single galaxies, as well as galaxy groups and
clusters. For example, feedback is thought to be key in determining
the upper mass cut-off of the galaxy mass function \citep[][]{benson,
  croton05, bower06}, balancing the radiative losses in cool-core
galaxy clusters, as well as providing the non-gravitational
`pre-heating' that may be important for non cool-core clusters
\citep[e.g.][]{bab,mac04,mc08}. Feedback also seems to play a role in
determining the relationship between the temperature and X-ray
luminosity of hot atmospheres across a range of halo masses
\citep[][]{bab,mac04,puchwein, bower08, dave, pope09}. As a direct
consequence, feedback also regulates supermassive black hole growth
\citep[e.g.][] {sr,chur06} and, therefore, its relation to the
properties of the host galaxy.

Considerable effort has been invested in understanding the impact of
supermassive black holes. One of the key challenges is understanding
how the energy from the AGN affects and couples to the broader
environment. Most studies that have sought to describe this
relationship have focused on bulk motion, shock waves and
pressure/gravity waves induced by AGN outbursts and the subsequent
dissipation of the associated energy \citep[see][for a
  review]{mcnuls}. In this article we focus on the transport of
material out of the cluster centre by AGN-blown bubbles. As an
example, the filaments of cool material observed behind AGN-blown
bubbles \citep[e.g.][]{consel,crawf05,hatch} are a strong indication
of bubble-induced mass transport within several 10s of kiloparsecs of
the cluster centre. The most obvious example is the Perseus cluster in
which the filaments contain some $ 10^{8}$ solar masses of cool
material \citep[e.g.][]{salome,salome08}.  More generally, some of the
transport mechanisms we discuss in this paper may carry matter out to
$\sim$100 kiloparsecs from the cluster centre.

Mass transport by bubbles not only reduces the amount of material
available for forming new stars in the central galaxy, allowing the
normally adopted stringent requirement that AGN heating perfectly
balance cooling to be some relaxed, but also affects the bubble
dynamics and energetics. Hence, a better understanding of the main
mass transport mechanisms is essential for describing the energy
balance in the ICM. Here, we focus on three important processes by
which bubbles can transport material: drift, wake transport and the
entrainment of ambient material into the bubble.

The aim of this article is to quantitatively describe these three main
mechanisms and also the corresponding implications.  The discussion is
largely analytical, intended to facilitate a better understanding of
the underappreciated aspects of these mechanisms and aid in the
interpretation of both observations and numerical simulations.

The article is arranged in the following way.  Section 2 outlines the
basic bubble model that serves as a backdrop for subsequent
discussions.  Sections 3, 4 and 5 focus on each of the three
mechanisms --- drift, wake transport and entrainment.  In each
section, we discuss a process as well as associated, potentially
observable consequences, like trailing optical filaments.  In Section
6, we consider how mass transport modifies the ongoing debate about
whether or not AGN heating need balance cooling precisely.  We
summarize our key findings in Section 7.

\section{Bubble Model}

The magnitude and spatial extent of mass transport associated with
buoyant bubbles depend heavily on the dynamics and the energetics of
the bubble.  It is, therefore, essential to have a model which
accurately describes these processes.
 
In broad brushstrokes, an AGN-inflated bubble is believe to evolve as
follows: an AGN jet inflates a bubble of what is generally assumed to
be relativistic fluid in the intra-cluster medium. The details of the
inflation process itself are not well understood and models for the
bubble inflation range from nearly adiabatic to supersonic. In either
case, at the end of the inflation process, the bubble is expected to
be in pressure equilibrium with its surroundings.  As a result of the
mass density in the bubble being lower than that of the ambient
medium, the bubble is buoyant and will rise, moving away from the AGN
and in the case of galaxy clusters, the cluster centre, towards more
distant regions of lower density and pressure. The velocity of the
bubble is determined by the near balance between buoyancy and drag
forces with the ICM.  In response to dropping ambient pressure, the
bubble will expand.  The end stage of bubble evolution is not well
understood.  One possibility is that the bubble will be destroyed
during the rise by hydrodynamical (Raleigh-Taylor and
Kelvin-Helmholtz) instabilities; however, it is equally plausible that
ambient conditions (magnetic fields on the bubble surface or viscosity
in the ICM) suppress these instabilities, allowing the bubble to rise
to a radius where the bubble density becomes equal to that of its
environment, and buoyancy force vanishes.  In the latter case, the
bubble of relativistic fluid will hover at its equilibrium height as a
``ghost cavity'', invisible in the X-rays though it may still be
detectable in low frequency radio observations.

In the context of mass transport, the most important stage in the
evolution of the bubble is the second (or the rise) phase.  The
dynamics and thermodynamics of this particular phase have been
discussed widely
\citep[e.g.][]{chur02a,enss,circulation,nusser,nulsen2007}.
Typically, most analyses assume that the rise velocity of the bubble
is sufficiently slow such that the response of the medium internal to
the bubble to the bubble expansion is adiabatic.  This approximation
yields reasonable theoretical expectations.

In addition, most treatments also assume that the bubble density is
much less than that of the surrounding ICM and derive the relevant
relationships in the limit of an extreme density contrast.  For many
of the calculations, this approximation is probably adequate.
However, the density contrast determines the maximum height to which
the bubble rises, and so is critical for determining the magnitude and
the spatial scale of the mass transport phenomenon, as well as the
efficiency with which energy can be extracted from the bubble.  As we
will discuss in the following sections, mass transport processes endow
a bubble with an effective --- and in the case of entrainment, a real
--- finite density contrast.  A more complete description of bubble
dynamics and thermodynamics must therefore allow for a finite density
contrast.  At relevant sections of this paper, we will therefore
present modified relationships governing bubble dynamics and
thermodynamics that takes into account this aspect.

\section{Bubble Drift and Trailing Cool Filaments}

As a bubble rises, energy conservation requires that the change in its
gravitational potential energy is accompanied by a gain in kinetic
energy of the bubble and the ICM.  The latter is induced by the ICM
being pushed by the rising bubble, causing a net upward displacement
behind the bubble.  This phenomenon is known as ``drift''
\citep[][]{darwin} \footnote{The phenomenon of drift is utilised in
  industrial mixing processes, called fluidised beds, which are formed
  when gas is passed into a vessel containing a bed of solid
  particles. At a given gas injection rate, a stream of bubbles
  appears and rises through the bed with tails of material behind them
  and thereby intensifying mixing \citep[e.g.][]{crowe,yang}.}  Figure
1 depicts a schematic view of the bubble and its trailing drift
\citep[e.g.][]{crowe,yang,schetz}.

Drift is closely related to the better-known concept of ``added mass''
\citep[e.g.][]{nusser,pav}, a concept associated with moving bodies
submerged in a fluid.  The usual interpretation is that it represents
the increased inertia associated with the work required to change the
kinetic energy of the fluid flow around the body.  The ambient fluid
can be thought of as increasing the effective mass of the moving
object \citep[e.g.][]{milne}.

In group and cluster environments, this trailing fluid can, under
certain conditions, give rise to filaments of cool gas.  In this
section, we introduce the phenomenon of drift, discuss the associated
characteristics, and explore whether the recently discovered optical
filaments stretching between the AGN-inflated bubbles and the cluster
centres \citep[e.g.][]{consel,crawf05,salome,hatch} are related to the
drift.  The kinematic data \citep[e.g.][]{hatch} certainly suggests
that the filaments may well be the result of cool material that was
originally lifted from the cluster center and that some of this
material is now falling back.
  
 \subsection{Drift: a conceptual outline}

\cite{darwin} considered the problem of a body moving uniformly
through an infinite, inviscid, incompressible fluid. Importantly,
Darwin studied the motion in the fluid frame of reference rather than
the rest frame of the body in which the fluid particles follow simple
streamlines. In the fluid frame, the trajectories are extremely
complex, consisting of a large-scale looping motion and a permanent
displacement in the direction of body's motion.

The forward displacement occurs because a moving body pushes the fluid
ahead of it causing the fluid to move in the same direction as the
body. In due course, the forward-displaced fluid slows down, though in
the situation considered by \cite{darwin} never formally comes to rest
behind the body. As an illustration, one can imagine a rising bubble
approaching and traversing a stationary marked plane within the fluid
column, oriented at right angles to its direction of motion. As the
bubble passes through the plane, the fluid will be displaced upwards
in the horn-like shape depicted in the third panel figure 1, with the
fluid closer to the body experiencing a greater net translation.

Strictly speaking, the fluid displacement has two contributions:
$X_{\rm d}$, a localised (positive) drift contribution which decays
rapidly from the body, and $X_{\rm r}$, a non-local (negative) reflux
contribution which depends on the far-field boundary conditions and
which decays slowly from the body. In the limit that the body starts
infinitely far from the marked plane and then moves through the plane
to infinity, the reflux contribution is zero, so that the drift volume
is single-valued.

In the absence of gravity, the fluid displacement is permanent and the
length of the drift grows continuously while the total volume remains
approximately constant. However, for drift behind buoyant bubbles in
the intracluster environment, the stratification of the fluid cannot
be neglected. Not only will gravity bring the drifting fluid particles
to rest sooner, it can also potentially cause the fluid to fall back
to its original position.  As a result, in a stratified fluid, the
upward displacement of material due to drift is always finite.  The
implications of drift in a stratified fluid are described in detail in
the next subsection. Presently, we will assume, for simplicity's sake,
that the effects of gravity can be neglected.

The volume of fluid enclosed by the intial plane of marked fluid and
the distorted plane is called the drift volume, $V_{\rm drift}$, and
is depicted in figure 1. In an unbounded flow, it is calculated from
the integral of the permanent displacement of fluid particles in the
direction of the body's motion over the cross-sectional area of the
flow \citep[e.g.][]{eames}.

The displacement is given by the time integral of the relative
velocity between the fluid in the drift and the fluid at infinity. It
is calculated from the velocity potential for a uniform flow of an
incompressible, non-rotational fluid past a sphere,
e.g. \cite{lifschitz} \citep[see also][for recent
  examples]{pav,dursi}. Using this, \cite{darwin} showed that the
drift and body volumes are related by a constant of proportionality,
$k$, which depends on the geometry of the body; for a solid sphere, $k
= 0.5$. In the case where bubbles deform completely into vortex rings,
\cite{dabiri} found $k = 0.72$, so the value of $k$ for an ordinary
bubble should be somewhere between these limiting cases. For
generality, we will use \citep[e.g.][]{milne,schetz}
\begin{equation}\label{eq:15} 
V_{\rm drift} = k(V_{\rm bub}  +V_{\rm wake}),
\end{equation}
where $V_{\rm wake}$ is the wake volume. (For a full list of
parameters and constants, see the Appendix \ref{app:param}.)
\cite{darwin} was also able to show, quite generally, that the mass
within the drift volume is equal to the added mass, so that
\begin{equation}
M_{\rm drift} = k\rho_{\rm ICM} (V_{\rm bub} +V_{\rm wake})\approx k
M_{\rm dis},
\end{equation}
where $M_{\rm dis}$ is the ICM mass displaced by the bubble.   

At the onset of the rise phase, the (initial) displaced mass can be
estimated as $M_{\rm dis,0} = \rho_0 V_0$.  Given the enthalpy of the
bubble at the end of inflation and the beginning of the rise phase
\begin{equation}\label{eq:1}
E_{\rm bub,0} = \frac{P_0V_{\rm bub,0}}{\Gamma_{\rm b}-1} + P_0V_{\rm bub,0} =
\frac{\Gamma_{\rm b}}{\Gamma_{\rm b}-1}P_0V_{\rm bub,0},
\end{equation}                  
where $\Gamma_{\rm b}$ is the adiabatic index of the bubble
material, we can write 
\begin{equation} \label{eq:disp}
M_{\rm dis,0} = \frac{(\Gamma_{\rm b}-1)}{\Gamma_{\rm b}}\frac{\mu
  m_{\rm p}}{k_{\rm B}T_0} E_{\rm bub,0}\approx 10^{8}\bigg(\frac{E_{\rm bub,0}}{10^{59}\,{\rm
    erg}}\bigg)\bigg(\frac{T_0}{10^{7}\,{\rm K}}\bigg)^{-1}\,{\rm
  M_\odot}.
\end{equation}
Here, $T_0$ is the ambient temperature where the bubble is inflated
and we have assumed that the bubble contents are relativistic, hence
$\Gamma_{\rm b} = 4/3$.  We note that the actual value of $\Gamma_{\rm
  b}$ will depend on the initial composition of the AGN outflow and
whether the jet and the bubble inflation process entrains a
significant amount of material from the ICM into the bubble.  We will
comment on this further when we consider entrainment.

To estimate the drift mass that we would expect to be associated with
an AGN-inflated bubble, we assume that $k=0.5$ since the bubbles tend
to generally be more or less spherical.  Accordingly, we would expect
observations to show a total mass of a few $\times 10^{7}$ to $\sim
10^{8}\,{\rm M_\odot}$ of material trailing behind typical
bubbles. This value agrees reasonably well with estimates of the mass
in the recently discovered cool filaments that stretch between the
bubbles and the cluster centre \citep[e.g][]{salome08}.

In principle, it is possible to estimate $k$ for an AGN-blown bubble
directly from numerical simulations. Illustrative examples of the
characteristic drift signature in galaxy cluster simulations can be
found in \cite{revaz,roed}, though we note that this effect can also
be induced by circular vortex lines, which stretch bubbles into tori
\citep[e.g.][]{pav}. Furthermore, \cite{bub01} also estimated the mass
of material uplifted by their simulated hydrodynamic bubble to be
$\sim 10^{8}\,{\rm M_\odot}$.  This agreement is encouraging; however,
we do caution that until we achieve a better understanding of the ICM
microphysics (including a realistic treatment of magnetic fields,
thermal conduction and viscosity) any estimate for $k$ based on
numerical simulations is unlikely to significantly improve upon the
values quoted above. The same reasoning applies to three other
phenomenological parameters used throughout this article: (1) $q$, the
fraction of displaced mass trapped in the bubble wake; (2) $C_{\rm
  D}$, the bubble drag coefficient; (3) $\alpha$, the entrainment
coefficient. Plausible ranges and best estimates are given in table
\ref{tab:tab1}.

\begin{figure*}
\centering
\includegraphics[width=12cm]{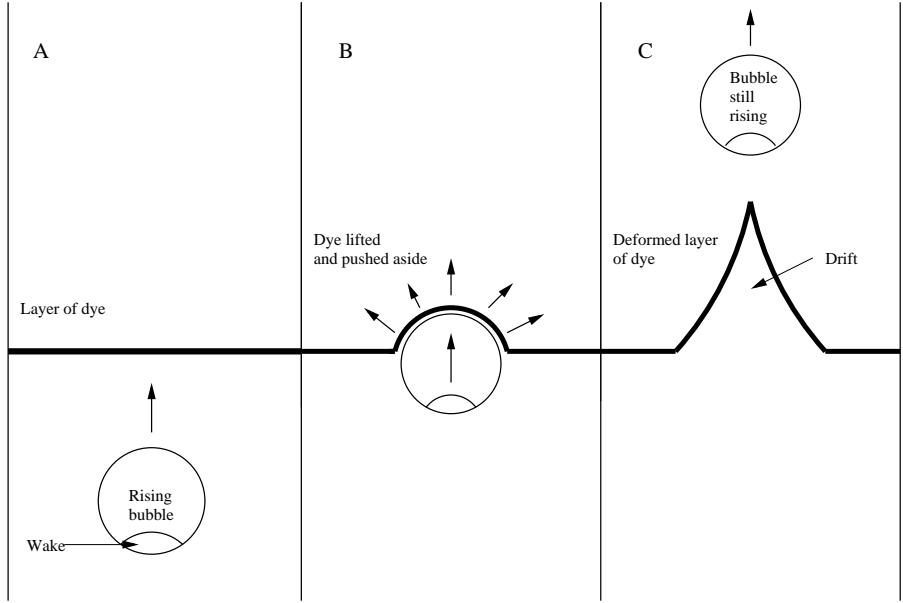}
\caption{Schematic showing wake transport and drift due to passage of
  bubble. In panel (A) the bubble approaches an initially planar dye
  surface from below. (B) Distortion of the dye surface occurs as the
  bubble passes through the plane. (C) The volume between initial
  plane and horn-like distorted surface is the drift volume.}
\label{fig:flow}
\end{figure*}

Finally, we note that drift is enhanced at low Reynolds numbers, as
can be seen in the viscous hydrodynamic simulations presented by
\cite{reynolds04}. However the theory is much more complex as the
drift volume does not have a well-defined maximum as well as being
dependent on boundaries in the system \citep[see][]{eames03}.
\begin{table}
\caption{Typical values of phenomenological constants}
\begin{tabular}{p{1in}p{1in}p{1in}}
  \hline
  Parameter & Plausible Range & Best guess\\
  \hline
  k & 0-1 & $\sim$ 0.5 \\
  q & 0-1 & $\sim$ 0.24 \\
  $C_{\rm D}$ & $>$ 0.45 & $>$ 0.45 \\
  $\alpha$ & 0-1 & $\lesssim$ 0.05\\
  \hline
\end{tabular}
\label{tab:tab1}
\end{table}

\subsection{Observed cool optical filaments: a by-product of the drift?}

Optical filaments behind bubbles tend to be linear trails of cool
material that extend for tens of kiloparsecs towards the cluster
centre  \citep[e.g.][]{consel,crawf05,salome,hatch}.  These structures
are a strong indication of bubble-induced mass transport within several 
10s of kiloparsecs of the cluster centre.  The mass associated with the 
filaments is estimated, in Perseus cluster at least, to be $\sim 10^{8}$ 
solar masses of cool material \citep[e.g.][]{salome,salome08}.

The simplest explanation for the filaments is that AGN-blown bubbles
lift cool material from the cluster centre when they detach and start
to rise.  The filaments would appear to grow when some of this
material decouples from the bubble-drift system and falls back
inward. Filamentary material close to the bubble would, in this
picture, still be travelling upward with respect to the cluster
centre, while material further away would be travelling towards the
cluster centre. The filament would then appear to be stretching, as
indicated by observations \citep[e.g.][]{hatch}.

The observable properties of filaments correspond quite closely to the
general characteristics of drift. We, therefore, investigate the
circumstances under which these processes may produce flows that might
be identifiable as filaments. Further to this, we also investigate
additional processes which may explain the clumpiness of filaments on
scales of tenths of kiloparsecs.

As previously mentioned, the stratification of the ICM exerts a
non-negligible influence on motions generated within it. The numerous
different flow regimes can be described by using the characteristic
length scales of the system. Pertaining to drift behind buoyant
bubbles, two useful scales are the Ozmidov and buoyancy lengths. The
first parameter relates to the maximum size of turbulent eddies in a
stratified fluid. Motions on larger spatial scales, therefore, cannot
be turbulent alone and material will also be transported vertically in
the fluid. Consequently, the Ozmidov length marks a characteristic
scale of fluid motion above which a filament may form behind a buoyant
bubble.

The scale length of vertical fluid motion is given by the buoyancy
length, which represents the complete conversion of vertical kinetic
energy to potential energy in a stratified fluid. It can be seen that
filaments will be formed if the buoyancy length exceeds the Ozmidov
length. These scales are considered in more detail below.

To a reasonable approximation, the Ozmidov scale can be calculated in
the following manner: the kinetic energy per unit mass required to
displace a stably stratified fluid a vertical distance $\Delta z$ is
of order $\Delta z^{2}N^{2}$, where $N$ is the Brunt-V\"ais\"al\"a
(buoyancy) frequency, while the energy associated with turbulence on
the same spatial scale is $(\epsilon \Delta z)^{2/3}$, with $\epsilon$
being the energy dissipation rate per unit mass.  Equating the kinetic
energies gives the Ozmidov scale \citep[e.g.][]{kantha}
\begin{equation}
L_{\rm O} = \bigg(\frac{\epsilon}{N^{3}}\bigg)^{\frac{1}{2}}.
\end{equation}
where
\begin{equation}
N^{2} = -\frac{g}{z}\frac{{\rm d}\ln \sigma}{{\rm d}\ln z},
\end{equation}
with $g$ being the gravitational acceleration and $\sigma = P
\rho_{\rm ICM}^{-\Gamma_{\rm e}}$ being the entropy index of the
ICM. It follows that, in a stratified fluid, there is a maximum
kinetic energy associated with the eddies that occur as a consequence
of a given body's motion. Therefore, for objects that generate
vertical disturbances on scales larger than $L_{\rm O}$, the turbulent
eddies cannot dissipate the kinetic energy. Instead, the majority of
the kinetic energy goes towards increasing the potential energy of the
ICM by carrying denser material from closer to the cluster centre out
to larger distances. Motion on much greater scales than $L_{\rm O}$,
is expected to generate gravity waves \citep[c.f.][]{omma04}. In
contrast, for fluid disturbances that are smaller than $L_{\rm O}$,
turbulent eddies are highly efficient at disrupting the column of
material displaced by drift, thereby preventing an appreciable change
in the potential energy of the fluid - in this case no filament will
form.

The Ozmidov scale, therefore, marks the transition from motions that
primarily generate turbulence to those that transport mass
vertically. Since turbulence is associated with heating, the
transition is equivalent to defining a threshold above which buoyant
bubbles predominantly transport mass and below which they generate
turbulence, which leads to heating. At this point it is also important
to note that if the Ozmidov scale becomes comparable to the size of
the smallest eddies, the Kolmogorov scale \citep[e.g.][]{plasmas},
then turbulence cannot exist in the fluid. This is because the eddy
kinetic energy is insufficient to circulate in the direction of the
fluid stratification. However, unless the ICM viscosity exceeds the
Spitzer value this is unlikely to be the case in clusters of galaxies.

To simplify the expression for $L_{\rm O}$, the turbulent dissipation
rate for the bubble is taken to be $\epsilon = c_{\rm turb} w^{3}/r$
\citep[e.g.][]{denchand}, where $c_{\rm turb}$ is a numerical
constant, $w$ is the bubble velocity, and $r$ is the bubble radius. In
order to describe the scaling properties of the Ozmidov length we will
assume that the bubble ascends at its terminal velocity governed by
the balance between buoyancy and drag, with a magnitude given by
\citep[e.g.][]{bub01}
\begin{equation} \label{eq:17}
w_{\rm terminal} = \bigg(\frac{8 r g}{3 C_{\rm
    D}}\bigg)^{\frac{1}{2}},
\end{equation}
where $C_{\rm D}$ is the drag coefficient. For brevity, equation
(\ref{eq:17}) is written $w_{\rm terminal} = (c_{\rm term} r
g)^{1/2}$ and the Ozmidov scale can be expressed as
\begin{equation}\label{eq:18}
L_{\rm O} \approx c_{\rm turb}^{\frac{1}{2}}c_{\rm term}^{\frac{3}{4}}
r^{\frac{1}{4}}z^{\frac{3}{4}}\bigg(\frac{{\rm d} \ln \sigma}{{\rm
    d}\ln z}\bigg)^{-\frac{3}{4}},
\end{equation}
where $z$ is the bubble's displacement from the cluster centre.

Equation (\ref{eq:18}) must be compared to the buoyancy length to
determine whether filaments actually form. The buoyancy length is
defined as the displacement of a fluid parcel at which all its
vertical kinetic energy is converted to potential energy. As before,
the kinetic energy per unit mass, $w^{2}$, will displace fluid in a
stably stratified atmosphere by a vertical distance $L_{\rm buoy}$, is of
order $L_{\rm buoy}^{2}N^{2}$, so that \citep[e.g.][]{kantha} the
buoyancy length is
\begin{equation}
L_{\rm buoy} = \frac{w}{N}.
\end{equation}
For a bubble travelling at its terminal velocity $w = (c_{\rm term} r
g)^{1/2}$ \citep[e.g.][]{bub01}, and
\begin{equation}\label{eq:19}
L_{\rm buoy} \approx c_{\rm term}^{\frac{1}{2}}
r^{\frac{1}{2}}z^{\frac{1}{2}}\bigg(\frac{{\rm d} \ln \sigma}{{\rm
    d}\ln z}\bigg)^{-\frac{1}{2}}.
\end{equation}
Following the discussion in \cite{denchand}, our constant $c_{\rm
  turb} = 0.42/2 = 0.21$. Then, using the fact that, for a spherical
bubble $c_{\rm term} = 8/(3C_{\rm D})$, it can be shown that filaments
can form behind buoyant bubbles provided
\begin{equation}\label{eq:20}
\frac{L_{\rm buoy}}{L_{\rm O}} \approx 1.7 C_{\rm D}^{\frac{1}{4}}
\bigg(\frac{r}{z}\bigg)^{\frac{1}{4}}\bigg(\frac{{\rm d}\ln
  \sigma}{{\rm d}\ln z}\bigg)^{\frac{1}{4}} > 1.
\end{equation}
Whether the criterion above is satisfied depends on the values given
to the phenomenological constants, $c_{\rm term}$ and $C_{\rm D}$, as
well as observable parameters. It is particularly difficult to assign
a characteristic value for the drag coefficient, $C_{\rm D}$, since it
depends on both the Mach and Reynolds numbers of the flow
\citep[e.g. see][and references therein]{pope08a}, which are dependent
on the nature of the ICM. Estimating such parameters by comparison
with numerical hydrodynamic simulations is not necessarily
representative. However, a suitable guess can be obtained in the limit
that $C_{\rm D}$ behaves as a conventional drag coefficient so that it
varies from $C_{\rm D} \approx 30$ for a Reynolds number of $Re
\approx 1$, to $C_{\rm D} \approx 0.5$ for $Re > 1000$
\citep[e.g.][]{multi}. This suggests that filaments are more easily
generated in flows with a low effective Reynolds number, which may be
the case for the rarefied ICM. Despite these uncertainties, equation
(\ref{eq:20}) remains informative because it indicates the functional
relationship between physical parameters.

The observational quantities are slightly better constrained. In
classical cool-core clusters, the entropy profile scales roughly
linearly with distance all the way to the cluster centre
\citep[e.g.][]{son09}, with a logarithmic gradient close to unity. We
also assume the empirical relation that the bubble radius is linearly
proportional to its displacement from the cluster centre, such that
$r/z \sim 0.3-0.5$ \citep[e.g.][]{diehl28,mcnuls}. Combining these
values gives $L_{\rm buoy}/L_{\rm O} \sim 1-3$, indicating that
material can potentially be displaced upwardly behind the bubble. Some
of this material will come from the centre of the cluster and will be
cool, meaning that an optical filament could form behind the rising
bubble.  Interestingly, \cite{son09} present entropy profiles for a
sample clusters, of which at least A1795, A2199 and A478 are nearly
linear entropy profiles and also show evidence for filamentary
structures \citep[see][and references therein]{pope08a}.  We also note
that the typical mass in observed filaments of $\sim 10^{8}\,{\rm
  M_\odot}$ \citep[e.g][]{salome08} agrees well with drift mass
estimates for typical ICM bubbles.  Additionally, the closeness of
$L_{\rm buoy}/L_{\rm O}$ to unity in clusters with nearly linear
entropy profiles also suggests that buoyant bubbles in these systems
do not generate gravity waves, though powerful long-lived episodes of
jet activity might \citep[see, for example][]{omma04}.

Classical cool core clusters, however, comprise only a small fraction
of the clusters. In general, cluster entropy profiles tend to flatten
close to the centre and then increase roughly linearly with distance
from the cluster centre \citep[e.g.][]{don06}.  If the entropy profile
has a flat core when the bubble begins to rise (as opposed to being
initially nearly linear and subsequently evolving into a cored profile
due to the resulting heat injection), the logarithmic entropy gradient
is approximately zero close to the cluster centre (see equation
\ref{eq:18}) and then roughly constant with a value near unity further
out, e.g. 1.0-1.3 \citep[e.g.][]{don06,cav09}. In this case, $L_{\rm
  buoy}/L_{\rm O}$ will be small near the cluster centre and the
bubble motion will primarily generate turbulence which heats the
central regions. Outside the central core, $L_{\rm buoy}/L_{\rm O}$
will take a value similar to that quoted above for the classical
cool-core clusters and some material may be displaced upwards by the
bubble. However, since the upward displacement can only occur outside
the central entropy core, there is unlikely to be a significant amount
of cool material in the drift region. In this case, the drift will
likely not appear as optical filaments.

\section{Wake Transport}

Apart from the drift, a rising bubble can also transport material via
``wake transport''.  When a deformable bubble starts to rise, it takes
on a shape that is roughly spherical, but with an indentation at the
bottom.  This cavity between a true sphere and the actual bubble shape
with an indented base is known as the wake
\citep[e.g.][]{crowe,yang,schetz}.  Figure 1 depicts a schematic view
of the bubble, wake and drift. The ``wake'' in the present context is
very different from the the turbulent wake that develops behind
objects at high Reynolds numbers.

Studies of bubbles and mass transport by bubbles in non-cosmological
settings indicate that line vortices generated behind the bubble as it
begin to rise lift, trap the ambient medium in the vicinity
(i.e. material in the cluster centre) within the ``wake'' cavity
\citep[c.f.][]{pav}.  This trapped material is then carried up by the
bubble.

In the first instance, material in the wake increases the average
density of the bubble and as a direct consequence reduces its ascent
velocity and the maximum height to which the bubble rises.  In
addition, trapped material can also drop out of the wake, and this
material will further contribute to the formation of filaments behind
AGN-blown bubbles.

\subsection{Wake: a conceptual outline}

Studies of fluidised beds indicate that a typical wake contains a
fraction $q$ of the initial mass displaced by the bubble:
\begin{equation} \label{eq:wake}
M_{\rm wake}\approx q M_{\rm dis,0}, 
\end{equation}
where $q \approx 0.24$ and $M_{\rm dis,0}$ is given by equation
(\ref{eq:disp}).  Assuming that this laboratory value of $q$ is
applicable to the intra-cluster medium, the above relationship
suggests that AGN-inflated can easily lift few $\times 10^{7}$ to
$\sim 10^{8}\,{\rm M_\odot}$ in their wakes.

In addition to facilitating the transport of material out of the
cluster centre, the wake --- and specifically, the mass in the wake
--- increases the effective average density of the bubble (though it
does not affect the bubble's actual density contrast as measured from
X-rays).  As we discuss below, this modification impacts not only the
maximum height that the bubble can ascend to, but also the efficiency
with which energy is extracted from it.

We do, however, note that as in the case of the geometric constant $k$
in the estimate for the drift mass, we acknowledge an uncertainty: it
is unclear whether the laboratory value of $q$ is applicable to the
intra-cluster medium.  It is difficult to propose a better motivated
value because our understanding of the ICM microphysics is highly
incomplete.  This also means that the estimates of $q$ based on
numerical simulations are not necessarily more believable than the
laboratory value deduced from the observations of fluidised beds.  On
the other hand, one would expect that $q$ does not exceed unity, since
this would imply that the wake contains more mass than originally
displaced by the bubble; consequently, unless $q \ll 1$ for the ICM,
our estimates for the wake mass in the range of $\sim 10^{8}\,{\rm
  M_\odot}$ should be reasonable.

\subsection{The implications of wake transport}\label{wake:b}

Studies of bubbles indicate that during the ascent, the bubble wake
can lose blobs of material \citep[][]{crowe}.  Formally, this happens
if the terminal infall velocity of a blob exceeds the local upward
velocity behind the bubble.  Once detached from the wake, the
subsequent evolution of the material is unclear.  In the simplest
picture, the blob simple falls back inward.  The presence of the ICM
complicates the scenario in that the associated ram pressure will
likely strip and stretch the blob into filamentary structures.
Moreover, the material from the shredded blobs may also get caught in
the drift behind the bubble and some of this material may still get
displaced to larger distances from the cluster centre
\citep[][]{crowe,yang}.

This discussion would suggest that both drift and leakage of material
from the bubble wake can contribute to the formation of trailing
filaments.  However, the bulk of the wake material will tend to fall
towards the cluster centre.  As a result, the trailing material is
unlikely to exhibit a single characteristic kinematic signature.
Rather, the kinematic structure will depend on the relative importance
of drift and leakage from the wake, and will probably vary
significantly over time as different processes become prevalent. This
issue bears further investigation.

As already noted, both the wake and drift affect the dynamics of the
bubble, but in slightly different ways.  To illustrate this, the
equation of motion of a bubble carrying a wake, and subject to the
action of drift, can be written as
\begin{align}\label{eq:21}
&\frac{{\rm d}[(M_{\rm bub} +  M_{\rm wake} +  M_{\rm drift})w]}{{\rm d}t} \\ \nonumber
& \qquad\qquad\qquad\qquad =  g(M_{\rm bub}+M_{\rm wake} - M_{\rm dis}) - F,
\end{align}
where $M_{\rm dis}$ is the mass displaced by the bubble, which falls
with time, and $F$ is the drag force
\begin{equation}\label{eq:22}
F = \frac{1}{2}C_{\rm D}A \rho_{\rm ICM}w^{2}
\end{equation}
and $A=\pi r^{2}$ is the cross-sectional area of the spherical bubble.

From the left hand side of equation (\ref{eq:21}) it can be seen that
a bubble of constant mass, $M_{\rm bub}$, which transports the
additional masses, $M_{\rm wake}+M_{\rm drift}$, accelerates as if it
has a larger effective mass of $M_{\rm bub}+M_{\rm wake}+M_{\rm
  drift}$. However, in the limit that the acceleration is zero, the
bubble behaves slightly differently again; the terminal velocity of
the bubble described by equation (\ref{eq:21}), can be expressed as
\begin{equation}\label{eq:22}
w_{\rm terminal} = \bigg[\frac{8}{3}\frac{gr}{C_{\rm D}}\bigg(1 -
  \frac{M_{\rm wake}+M_{\rm bub}}{M_{\rm
      dis}}\bigg)\bigg]^{\frac{1}{2}}.
\end{equation}
Equation (\ref{eq:22}) can be interpreted as the terminal velocity of
a bubble with an effective density contrast of
\begin{equation}\label{eq:23}
\eta_{\rm eff} = \frac{M_{\rm dis}}{M_{\rm wake}+M_{\rm bub}}.
\end{equation}

Denoting the effective density contrast when the bubble first starts
to rise after inflation as $\eta_{\rm eff,0}$, we note that
\begin{equation}\label{eq:eta}
\eta_{\rm eff,0} = \frac{\eta_0}{q\eta_0 +1},
\end{equation}
where $\eta_0 \equiv \rho_{0}/\rho_{\rm bub,0} = M_{\rm dis,0}/M_{\rm
  bub}$ is the {\em actual} initial density contrast and $M_{\rm
  dis,0}=\rho_{0}V_0$ is the initial displaced mass.  Here, $\rho_{0}$
is the ICM density where the bubble is inflated, and $\rho_{\rm
  bub,0}$, $V_0$ are the bubble mass density and the bubble volume
once the bubble is inflated.  A typical value of the actual initial
density contrast taken from the literature is $\eta_0 \sim 100$
\citep[e.g.][]{ghizzardi04}.  With mass in the wake characterized by
$q \approx 0.24$ \citep[see][]{yang}, $\eta_{\rm eff,0} \approx 4$.
In this derivation, we have assumed that the mass of the bubble
remains constant during the rise.  This is equivalent to assuming that
regardless of whether the ICM is entrained within the bubble at
formation, the bubble does will not subsequently entrain any
additional material during the rise.  We have also assumed that the
wake mass remains constant.  The displaced mass is allowed to change
since both the bubble volume and the ambient density vary with
distance from the cluster centre and for completeness, we note that
this mass will decrease as the bubble rises.

According to equation (\ref{eq:22}), the low effective density
contrast reduces the terminal velocity of the bubble.  In the initial
stages, $\eta_{\rm eff,0}\approx 4$, results in a lower terminal
velocity by a factor of $\sim 0.9$.  And even though the effect is not
large, it is worth bearing in mind since the bubble rise timescale, a
quantity commonly used to estimate AGN power, is inversely
proportionate to the velocity.

A lower density contrast also reduces the equilibrium height to which
the bubble can rise and a sufficiently low contrast can result in the
bubble not being able to rise beyond the very central regions of the
cluster.  This can be illustrated as follows.

Under the simplest circumstances, and ignoring fluid instabilities,
the buoyant bubble will rise through the stratified atmosphere and
expand adiabatically to maintain pressure equilibrium with its
surroundings.  Consequently, the bubble density can be related to the
ambient density and temperature by \citep[see also][]{hint07}
\begin{equation}\label{eq:3}
\frac{\rho_{\rm bub}}{\rho_{\rm bub,0}} = \bigg(\frac{\rho_{\rm
    ICM}}{\rho_0}\bigg)^{\frac{1}{\Gamma_{\rm b}}}\bigg(\frac{T_{\rm
    ICM}}{T_0}\bigg)^{\frac{1}{\Gamma_{\rm b}}},
\end{equation}
where $T_0$ and $\rho_0$ are the ICM temperature and density,
respectively, where the bubble was inflated, $T_{\rm ICM}$ and
$\rho_{\rm ICM}$ are the ambient temperature and density at the
bubble's current location.

This relationship shows that the bubble density drops more slowly than
the ambient density and there will be a location where the densities
become comparable.  In the absence of a wake, the buoyancy force would
also vanish at this height, and the upward motion of the bubble would
stagnate since this motion is driven by buoyancy.  In practice, due to
the kinetic energy gained by the bubble, it will not immediately come
to halt at the equilibrium height.  Instead, it will overshoot this
location.  Above the equilibrium height, the bubble will be denser
than its surroundings and experience a gravitational restoring force
towards the cluster centre. Consequently, it will oscillate around the
equilibrium radius at approximately the Brunt-V\"ais\"al\"a (buoyancy)
frequency \citep[e.g.][]{plasmas}.  The oscillation will, in due
course, be damped by the drag force but while it is oscillating, the
bubble will generate fluid disturbances that travel towards and away
from the cluster centre.  Consequently, a fraction of the kinetic
energy gained during the ascent can go towards heating the central
regions of the ICM.

The arresting criterion for the bubble-wake system, in the presence of
the wake, is obtained from equation (\ref{eq:21}) by noting that at
the equilibrium height, $z_{\rm eq}$
\begin{equation}\label{eq:24} 
M_{\rm wake}(z_{\rm eq}) \approx M_{\rm dis}(z_{\rm eq})- M_{\rm bub},
\end{equation}
where $M_{\rm dis}(z_{\rm eq})=\rho_{\rm ICM}(z_{\rm eq}) V_{\rm bub}
(z_{\rm eq})$ is the displaced mass at the equilibrium location.  This
criterion is equivalent to $\eta_{\rm eff} (z_{\rm eq}) = 1$.
  
Since $\eta_{\rm eff}$ can be written as 
\begin{equation}\label{eta-eff} 
\eta_{\rm eff}=\eta_{\rm eff,0}  \bigg(\frac{\rho_{\rm
    ICM}}{\rho_0}\bigg)^{\frac{\Gamma_{\rm b}-1}{\Gamma_{\rm b}}}\bigg(\frac{T_{\rm
    ICM}}{T_0}\bigg)^{\frac{-1}{\Gamma_{\rm b}}},
\end{equation}
then $\eta_{\rm eff} (z_{\rm eq}) = 1$ at the equilibrium height implies,
\begin{equation}\label{eq:4}                                              
\frac{\rho_{\rm ICM}(z_{\rm eq})}{\rho_0} =
\eta_{\rm eff,0}^{-\frac{\Gamma_{\rm b}}{\Gamma_{\rm b}-1}}\bigg[\frac{T_{\rm
    ICM}(z_{\rm eq})}{T_0}\bigg]^{\frac{1}{\Gamma_{\rm b}-1}}.
\end{equation}                                                        
where $T_{\rm ICM}(z_{\rm eq})$ is the ICM temperature at the
equilibrium height.  For cool-core clusters we typically expect $T_{\rm
  ICM}(z_{\rm eq})/T_0 \sim 3$ \citep[e.g.][]{allen01}.

Modelling the ICM density distribution using a $\beta$-profile
\begin{equation}\label{eq:5} 
\rho_{\rm ICM}(z) = \rho_{\rm ICM,0}{\bigg[1 + \left(
    \frac{z}{z_{\rm ICM,0}}\right)^{2}\bigg]^{-\beta}},
\end{equation}
where $\rho_{\rm ICM,0}$ is the central ICM density, $z_{\rm ICM,0}$
is the ICM scale height, and $z$ is the distance from the cluster
centre, we can estimate the equilibrium height to which an adiabatic
buoyant bubble carrying a wake, inflated at height $z_0$, will rise in
such an atmosphere:
\begin{align} \label{eq:6}                                           
 \bigg(\frac{z_{\rm eq}}{z_0}\bigg)^{2} & = \left[
   1+\left(\frac{z_{\rm ICM,0}}{z_{0}}\right)^2\right] \\ \nonumber &
 \qquad\qquad\times \eta_{\rm eff, 0}^{\frac{\Gamma_{\rm
       b}}{\beta(\Gamma_{\rm b}-1)}}\bigg[\frac{T_{\rm ICM}(z_{\rm
       eq})}{T_0}\bigg]^{-\frac{1}{\beta(\Gamma_{\rm b}-1)}} -
 \left(\frac{z_{\rm ICM,0}}{z_0}\right)^2
\end{align}

Restricting ourselves to classical cool core clusters that generically
exhibit very small scale heights and hence, nearly power-law density
profiles, we find that
\begin{equation} \label{eq:6a}                                           
 \bigg(\frac{z_{\rm eq}}{z_0}\bigg)^{2} \approx \eta_{\rm eff, 0}^{\frac{\Gamma_{\rm
      b}}{\beta(\Gamma_{\rm b}-1)}}\bigg[\frac{T_{\rm ICM}(z_{\rm
    eq})}{T_0}\bigg]^{-\frac{1}{\beta(\Gamma_{\rm b}-1)}} 
\end{equation}

In Figure 2, we plot equation (\ref{eq:6a}) as a function of $q\equiv
M_{\rm wake}/M_{\rm dis,0}$ for three different values of the initial
density contrast $\eta_{\rm eff,0} = 1000, 100, 10, 4$, assuming
$\Gamma_{\rm b}=4/3$, $\beta=3/4$ and $T_{\rm ICM}(z_{\rm
  eq})/T_{0}=3$ (as is appropriate in cool core clusters).

It can be seen that a low effective density contrast due to the wake
can significantly affect the equilibrium height, resulting in the
bubble being trapped closer to the cluster centre. Specifically, we
find that if we ignore the wake (i.e.~set $q=0$), then $\eta_{\rm eff,
  0}=\eta_{0}\approx 100$ and $z_{\rm eq}\sim {\rm few} \times 10^3
z_0$, which probably greatly exceeds the cluster virial radius even if
$z_0$ is small (i.e.~$z_0\approx 1$---$30\,$kpc). For the same
situation, if $q=0.24$ and $\eta_0=100$, then $\eta_{\rm eff,0}\approx
4$ and $z_{\rm eq}\sim 4.5 z_0$. If the bubble is inflated and begins
its ascent from $z_0=10\,$kpc, it will arrest at $\sim 50$ kpc.

\begin{figure*}
\centering
\includegraphics[width=10cm]{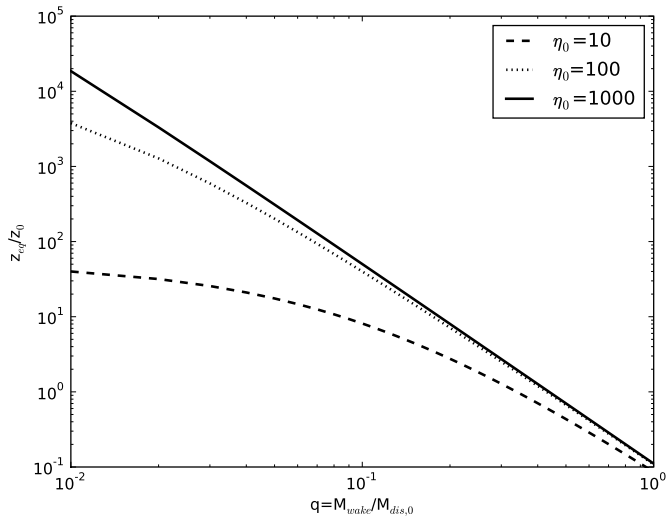}
\caption{Plot showing the equilibrium height as a function of $q =
  M_{\rm wake}/M_{\rm dis,0}$, for three values of $\eta_0$ using
  equation (\ref{eq:6a}). For this plot, $\Gamma_{\rm b} = 4/3$,
  $\beta=3/4$ and $T_{\rm ICM}(z_{\rm eq})/T_0 =3$.}
\label{fig:flow}
\end{figure*}

For completeness, we note that the ratio of the final to initial
pressure is
\begin{equation}\label{eq:7} 
\frac{P(z_{\rm eq})}{P_0} = \bigg[\eta_{\rm eff,0}\frac{T_0}{T_{\rm
      ICM}(z_{\rm eq})}\bigg]^{-\frac{\Gamma_{\rm b}}{\Gamma_{\rm
      b}-1}}.
\end{equation}

It can also be shown that the bubble's density contrast is important
for governing the efficiency with which energy is extracted. The total
energy available for heating the ICM and transporting mass is obtained
by integrating the buoyancy force acting on the bubble along its path
of motion \citep[c.f.][]{calori, nusser}
\begin{align}\label{eq:9}                                             
\int_{z_{\rm 0}}^{z_{\rm eq}} F_{\rm buoy}{\rm d}z = & \int_{z_{\rm
    0}}^{z_{\rm eq}} V_{\rm bub}\rho_{\rm bub}g {\rm d}z \\ \nonumber
& - \int_{z_{\rm 0}}^{z_{\rm eq}} V_{\rm bub} \rho_{\rm ICM}g {\rm d}z
+ \int_{z_{\rm 0}}^{z_{\rm eq}} M_{\rm wake} g {\rm d}z
\end{align}
where $g$ is the gravitational acceleration. Making use of the fact
that the ICM is in hydrostatic equilibrium, the second term on the
right hand side of equation (\ref{eq:9}) can be re-written as
\begin{equation}  \label{eq:10}                                           
V_{\rm bub} \rho_{\rm ICM} g {\rm d}z \approx V_{\rm bub} \frac{{\rm
    d}P}{{\rm d}z}{\rm d}z = V_{\rm bub} {\rm d}P,
\end{equation}  
where $P$ is both the pressure in the bubble and the ICM pressure.
These two are assumed to be equal since the bubble is in pressure
equilibrium with its surroundings.  Using equation (\ref{eq:7}) and
the fact that the bubble expands adiabatically,
\begin{align}\label{eq:11}  
\int_{P_0}^{P(z_{\rm eq})} & V_{\rm bub} {\rm d}P = \frac{\Gamma_{\rm
    b}}{(\Gamma_{\rm b}-1)}P_0V_0 \bigg[ \left(\frac{T_{\rm
      ICM}(z_{\rm eq})}{\eta_{\rm eff,0}T_0}\right) -1 \bigg],
\end{align}
where we have used equation (\ref{eq:7}) to derive the latter
relationship.

The remaining two terms on the right side of equation (\ref{eq:9}) can
also be approximated if we make some reasonable assumptions about the
bubble-wake-ICM system.  For example, for a bubble-wake system where
$M_{\rm bub}+M_{\rm wake}$ is constant, the requirement that the ICM
is in hydrostatic equilibrium allows us to write
\begin{align}\label{eq:11b}
\int_{z_{\rm 0}}^{z_{\rm eq}} V_{\rm bub}\rho_{\rm bub}\, g {\rm d}z +
& \int_{z_{\rm 0}}^{z_{\rm eq}} M_{\rm wake} \, g {\rm d}z
\\ \nonumber & \approx \frac{M_{\rm dis,0}}{\eta_{\rm eff,0}}
\int_{P_{\rm 0}}^{P(z_{\rm eq})} \frac{{\rm d}P}{\rho_{\rm
    ICM}}.
\end{align}
To make further progress, we restrict ourselves to classical cool core
clusters whose entropy profiles scale nearly linearly with distance
from the cluster centre \citep[e.g.][]{son09} and in keeping the
observations, we approximate that the ICM density profile for
classical cool core clusters as a power-law: $\rho_{\rm ICM} \approx
\rho_0 \left( z/z_0\right)^{-2\beta}$.  This profile is equivalent to
the $\beta$-profile in equation (\ref{eq:5}) in the limit $z_0 \gg
z_{\rm ICM,0}$, where $z_0$ is the radius at which the bubble is
inflated.  For $\beta=3/4$, we find that
\begin{align}\label{eq:11c}
\int_{z_{\rm 0}}^{z_{\rm eq}} V_{\rm bub}\rho_{\rm bub}\, g {\rm d}z +
& \int_{z_{\rm 0}}^{z_{\rm eq}} M_{\rm wake} \, g {\rm d}z
\\ \nonumber & \approx \frac{\Gamma_{\rm b}}{(\Gamma_{\rm
    b}-1)}\frac{P_0V_0}{\eta_{\rm eff,0}} \ln\left(\frac{T_{\rm
    ICM}(z_{\rm eq})}{\eta_{\rm eff,0}T_0}\right),
\end{align}
where we have applied equation (\ref{eq:7}).

Combining equations (\ref{eq:11}) and (\ref{eq:11c}), we find that
\begin{align}\label{eq:11d}                                             
 \int_{z_{\rm 0}}^{z_{\rm eq}} F_{\rm buoy}& {\rm d}z \approx
 \frac{\Gamma_{\rm b}}{(\Gamma_{\rm b}-1)}P_0V_0 \\ \nonumber & \times
 \left[ 1 + \frac{1}{\eta_{\rm eff,0}} \ln\left(\frac{T_{\rm
       ICM}(z_{\rm eq})}{\eta_{\rm eff,0}T_0}\right) -
   \left(\frac{T_{\rm ICM}(z_{\rm eq})}{\eta_{\rm
       eff,0}T_0}\right)\right].
\end{align}
For $T_{\rm ICM}(z_{\rm eq})/T_0 \approx 3$, comparing the energy
available for heating when $\eta_{\rm eff,0}\approx 100$ and
$\eta_{\rm eff,0}\approx 4$, which is the more appropriate value when
the wake mass is taken into account and corresponds to $q\approx
0.24$, we find that heating is reduced by a factor of $\sim 6$ in the
latter case.  Physically, this means that some of the bubble's energy
goes into raising the potential energy of its content, including the
wake, and is therefore unavailable for heating the ICM. The lower the
density contrast between the bubble and its surroundings, the lower
the fraction of energy released during the ascent.

\section{Entrainment}

Mass transport may also proceed by the entrainment of ICM material
into the bubble --- the drawing-in and incorporation of material from
the ICM into the bubble body --- and its subsequent displacement to
larger radii in the cluster, as the bubble rises \citep[c.f.][]{mc08}.

Entrainment can impact bubble dynamics in two important ways: (1) It
causes the bubble to expand more rapidly than predicted by a simple
adiabatic model, and (2) it limits the height to which a bubble can
rise.  The importance of this particular channel, however, depends on
the extent to which material is drawn into the bubble.

Establishing the extent of entrainment in AGN-inflated bubbles is a
non-trivial task.  Entrainment is an extremely complex phenomenon and
can occur in several ways.  For example, the ambient material may not
be displaced by the jet with 100\% efficiency, resulting in the
entrainment of some ICM material within the bubble.  Kelvin-Helmholtz
instabilities due to shear between the jet and the ICM can also draw
ambient material in through the sides of the jet.  Additionally,
Rayleigh-Taylor instabilities on the upper bubble surface can allow
the ICM material to penetrate, resulting in a time-varying mass
entrainment rate as the bubble rises.  On the other hand, given that
the observed bubbles appear to be highly resilient to instabilities
that can transform spherical bubbles into vortex rings, a number of
researchers have suggested that the AGN-blown bubbles may either be
draped by surface magnetic fields, or that the ICM may be more viscous
than thought \citep[e.g.][]{deyoung03,instab105}.  Both surface
magnetic fields and enhanced ICM viscosity dramatically inhibit mixing
and the growth of fluid instabilities.  But even in these situations,
intra-cluster material can still enter the bubble.  For example, in
the case of a bubble surface draped by surface magnetic fields, the
ICM can still leak into the bubble as a consequence of magnetic
reconnection events and diffusion across the bubble/ICM boundary,
\citep[see][]{pope2010}.

The upshot here is that entrainment of the ICM into AGN-blown bubbles
is not governed by hydrodynamic processes alone. As a result, findings
from current generation of numerical hydrodynamic (and even
magnetohydrodynamic) simulations are not directly applicable; these do
not, as of yet, adequately capture the necessary microphysical
processes.  This then argues in favour of treating the process in a
more general manner, as we attempt to do here, with explicit (rather
than implicit) assumptions about the nature of the fluid flow.

\subsection{Entrainment: a conceptual overview}

As a starting point, we turn to the conservation equations for mass,
momentum and thermal energy that are often applied to the study of
plumes in the terrestrial atmosphere \citep[e.g.][]{turner3}. In doing
so, we avoid having to make any implicit assumptions about the fluid
behaviour of the ICM, in contrast to standard hydrodynamic
simulations. For example, the flow of an inviscid, rarefied fluid
around an obstacle can have a low effective Reynolds number
\citep[e.g.][]{rare}, which cannot be accounted for in the continuous
fluid approximation. Furthermore, the Reynolds number range accessible
to hydrodynamic simulations is $Re \lesssim n^{2}$, where $n$ is the
number of computational cells across the width of the
flow. Consequently, coarser spatial resolution implies a larger
effective viscosity that may act to reduce the growth rate of
instabilities on the bubble surface, which affects the entrainment
rate. Instead, we model the effective Reynolds number of the flow,
affected by rarefication and viscosity, through the drag and
entrainment coefficients. In particular, the drag coefficient is
related to the Reynolds number \citep[e.g. see][for an approximate
  functional form]{pope08a} and hence the effective
viscosity. Furthermore, we calibrate the entrainment rate to match
mixing rates due to the growth of Rayleigh-Taylor instabilities in
hydrodynamic simulations of buoyant bubbles in cluster atmospheres
\citep[e.g.][]{bub01,pav08}. Applicability to the ICM is also extended
by accounting for the different adiabatic indices of the bubble and
the ICM.

Integrating the equations then yields the bubble radius, velocity and
density as it evolves in an external medium. Importantly, since the
bubble is assumed to be spherically symmetric, the integration can be
performed to very high precision so that numerical errors are
significantly diminished. It is also staightforward to ensure the
equations are entirely self-consistent and that the appropriate
quantities are conserved.

Nevertheless, certain assumptions are required to solve the
conservation equations given below; the most important being that the
bubble remains spherical, intact and in pressure equilibrium with its
surroundings. This is observationally justifiable; with the possible
exception of M87 which appears to exhibit a vortex ring structure
\citep[][]{m87rad}. The ICM is also taken to be hydrostatic and we
omit any effects due to radiative processes and thermal
conductivity. Other than this, the derivation does not make any
assumptions about the shape of the ICM density or temperature
profiles.These assumptions are similar to many numerical simulations
of bubbles and are permissible for the following reasons. Radiative
losses from the bubble depend on its particle content, which is
difficult to determine and might plausibly be small. Thermal
conduction of heat across the bubble/ICM must be strongly suppressed,
probably by magnetic fields, otherwise bubbles would rapidly evaporate
\citep[][]{pav08}.

To make the model as transparent as possible, we have employed the
well-known entrainment hypothesis which states that the velocity, $v$,
at which material joins the bubble is proportional to the ascent
velocity of the bubble, $w$ \citep[][]{turner3}. The constant of
proportionality, $\alpha$, is called the entrainment coefficient and
is defined by
\begin{equation}\label{eq:27} 
v = \alpha w.
\end{equation}
In a uniform medium, and in the Boussinesq limit, the typical
entrainment coefficient for a momentum-driven outflow is $\alpha \sim
0.05$ and $\sim 0.08$ for a buoyancy-driven flow
\citep[e.g.][]{turner2}. \cite{dey06} also describes how the
entrainment coefficient may vary with the density contrast, $\eta$,
and Mach number, $Ma$, such that $\alpha \sim 0.2 (Ma/0.5)^{-1}
(\eta/2)^{1/2}$. However, this functional form is representative of
mixing due to non-linear Kelvin-Helmholtz instabilites, which are
suppressed on the surface of a magnetised AGN-blown bubble. As a
result, we assume $\alpha$ to be a constant.

Following equation (\ref{eq:27}), the equation of mass conservation
for an entraining bubble is
\begin{equation} \label{eq:28}                                                 
\frac{{\rm d}M_{\rm bub}}{{\rm d}t} = \alpha a_{\rm bub}\rho_{\rm
  ICM}w,
\end{equation} 
where $a_{\rm b}$ is the surface area of the bubble. The relation is
best expressed in terms of the distance from the cluster centre, so,
assuming that $\frac{{\rm d}}{{\rm d}t} = w \frac{{\rm d}}{{\rm d}z}$
and the bubble is spherical, we can write
\begin{equation}\label{eq:29} 
\frac{{\rm d}M_{\rm bub}}{{\rm d}z} = 4\pi r^{2}\alpha \rho_{\rm ICM}.
\end{equation}
Substituting $M_{\rm bub} = \rho_{\rm bub}V_{\rm bub}$ into equation
(\ref{eq:29}) gives
\begin{equation}\label{eq:30} 
\frac{{\rm d}r}{{\rm d}z} = \alpha \frac{\rho_{\rm ICM}}{\rho_{\rm bub}}
-\frac{r}{3}\frac{1}{\rho_{\rm bub}}\frac{{\rm d}\rho_{\rm bub}}{{\rm
    d}z}.
\end{equation}
Using equation (\ref{eq:30}), forms the basis for calibrating $\alpha$
against numerical simulations of bubbles in cluster atmospheres. The
length scale over which entrainment dominates the mass of the bubble
can be defined as the distance, $L$, during which the bubble entrains
a mass comparable to the initial displaced mass. To first order, the
entrainment length is given by
\begin{equation}\label{eq:32} 
L \equiv \bigg(\frac{1}{M_{\rm dis,0}}\frac{{\rm d}M_{\rm bub}}{{\rm
    d}z}\bigg)^{-1} \sim 11 \bigg(\frac{r_0}{5\,{\rm
    kpc}}\bigg)\bigg(\frac{\alpha}{0.15}\bigg)^{-1}\, {\rm kpc}
\end{equation}
where $r_0$ is the initial bubble radius. \footnote{The critical value
  of $\alpha$ above which equation (\ref{eq:32}) strictly holds occurs
  for $L = z_{ICM,0}$, so that $\alpha_{\rm crit} =
  r_0/(3z_{ICM,0})$. For the current setup with an initial bubble size
  of $r_0 =$ 5 kpc, and a density scale height of $z_{ICM,0} =
  18.6\,{\rm kpc}$, we find $\alpha_{\rm crit} \approx 0.09$.}
Hydrodynamic simulations of bubbles in galaxy clusters show that
bubbles almost completely mix with their surroundings before they have
ascended a distance of roughly 2-3 initial bubble radii
\citep[e.g.][]{bub01,robinson04,gardini,pav08}. On this evidence, the
effective entrainment coefficient due to the development of
Rayleigh-Taylor instabilities must be $\alpha \sim 0.1-0.15$. The
effect of viscosity and magnetic fields are equivalent to a reduction
of $\alpha$.

To proceed further, it is necessary to consider the momentum and
thermal energy equations. The momentum equation is given by equation
(\ref{eq:21}) using $k\rho_{\rm ICM} V_{\rm bub} = M_{\rm drift}$ and
a constant wake mass of $q M_{\rm dis,0}$. The thermal energy equation
can be derived from the following statement: the addition of a mass of
gas, ${\rm d}M_{\rm bub}$, at temperature $T_{\rm ICM}$, and at
constant pressure, changes the internal energy of the bubble, $c_{\rm
  p,bub}T_{\rm bub}M_{\rm bub}$. In this description, $c_{\rm p,bub}$
is the specific heat capacity at constant pressure, per unit mass of
the bubble and $c_{\rm p,bub} = \Gamma_{\rm b}k_{\rm B}/[\mu m_{\rm
    p}(\Gamma_{\rm b}-1)]$, where $\Gamma_{\rm b}$ is the adiabatic
index of the bubble material. The thermal energy equation is then
\begin{eqnarray}\label{eq:33} 
c_{\rm p,ICM}T_{\rm ICM}{\rm d}M_{\rm bub} + c_{\rm p,bub}T_{\rm
  bub}M_{\rm bub} = \\ \nonumber (T_{\rm bub}+{\rm d}T_{\rm
  bub})(M_{\rm bub}+{\rm d}M_{\rm bub})c_{\rm p,bub},
\end{eqnarray}
where $c_{\rm p,ICM} = \Gamma_{\rm ICM}k_{\rm B}/[\mu m_{\rm
    p}(\Gamma_{\rm ICM}-1)]$, is the specific heat at constant
pressure per unit mass of the ambient gas.

Expanding the right hand side of equation (\ref{eq:33}) up to first
order terms and rearranging gives
\begin{equation}\label{eq:34} 
\frac{{\rm d}\rho_{\rm bub}}{\rho_{\rm bub}} = \frac{{\rm d}M_{\rm
    bub}}{M_{\rm bub}}\bigg[1 - \frac{\Gamma_{\rm ICM}(\Gamma_{\rm
      b}-1)}{\Gamma_{\rm b}(\Gamma_{\rm ICM}-1)}\frac{\rho_{\rm
      bub}}{\rho_{\rm ICM}}\bigg].
\end{equation}
The density of the bubble is also affected by adiabatic
expansion. Assuming the atmosphere is hydrostatic, the change in
density can be written in terms of the gravitational acceleration of
the cluster
\begin{equation}\label{eq:35} 
\frac{{\rm d}\rho_{\rm bub}}{\rho_{\rm bub}} = \frac{{\rm d}P}{\Gamma_{\rm
    b} P} = \frac{\mu m_{\rm p}g {\rm d}z}{\Gamma_{\rm b}k_{\rm
    B}T_{\rm ICM}}.
\end{equation}
However, because the bubble now contains a mixture of fluids with
different adiabatic indices, its expansion must be described in terms
of a single, time-dependent effective adiabatic index calculated in
the following way. Entraining a mass of material, $\Delta M$, at the
same pressure, is equivalent to entraining a volume $\Delta V = \Delta
M/\rho_{\rm ICM}$. This increases the internal energy of the bubble by
an amount $P\Delta V/(\Gamma_{\rm ICM}-1)$, after which the total must
be
\begin{equation}\label{eq:upgamma}
\frac{P(V_{\rm bub} +\Delta V)}{\Gamma -1} = P\bigg[\frac{V_{\rm
      bub}}{\Gamma_{\rm b}-1} + \frac{\Delta V}{\Gamma_{\rm
      ICM}-1}\bigg],
\end{equation}
where $\Gamma$ is the effective adiabatic index. Then, after the $i$th
timestep of the numerical integration, the updated value of $\Gamma$
must be
\begin{equation}\label{eq:upgammax}
\frac{1}{\Gamma_{\rm i+1}-1} = \frac{1}{(V_{\rm i} + \Delta V_{\rm
    i})}\bigg[\frac{V_{\rm i}}{\Gamma_{\rm i}-1} + \frac{\Delta V_{\rm
      i}}{\Gamma_{\rm ICM}-1}\bigg],
\end{equation}
where $\Delta V_{\rm i}$ is entrained volume during the $i$th
timestep, so that $V_{\rm i}+\Delta V_{\rm i}$ is the total bubble
volume after the timestep, $\Gamma_{\rm i}$ was the effective
adiabatic index at the beginning of the timestep and $\Gamma_{\rm
  ICM}$ is the adiabatic index of the ICM. In the subsequent equations
and numerical calculations we use $\Gamma_{\rm b} = \Gamma_{\rm i}$.

Equations (\ref{eq:30}), (\ref{eq:21}), (\ref{eq:34}) and
(\ref{eq:35}) can then be rearranged to yield simultaneous
differential equations for the bubble radius, density and velocity
\begin{equation}\label{eq:36} 
\frac{{\rm d}r}{{\rm d}z} = \alpha \frac{\Gamma_{\rm ICM}}{\Gamma_{\rm
    b}}\frac{(\Gamma_{\rm b}-1)}{(\Gamma_{\rm ICM}-1)} - \frac{\mu
  m_{\rm p}rg}{3\Gamma_{\rm b}k_{\rm B}T_{\rm ICM}},
\end{equation}
\begin{eqnarray}\label{eq:37} 
\frac{{\rm d}\rho_{\rm bub}}{{\rm d}z} =
\frac{3\alpha}{r}\bigg[\rho_{\rm ICM}-\frac{\Gamma_{\rm
      ICM}}{\Gamma_{\rm b}}\frac{(\Gamma_{\rm b}-1)}{(\Gamma_{\rm
      ICM}-1)}\rho_{\rm bub}\bigg]\\ \nonumber + \frac{\mu m_{\rm
    p}\rho_{\rm bub}g}{\Gamma_{\rm b}k_{\rm B}T_{\rm ICM}},
\end{eqnarray}
and
\begin{eqnarray}\label{eq:38} 
\frac{{\rm d}w}{{\rm d}z} = \frac{-3\alpha w \rho_{\rm
    ICM}/r}{\rho_{\rm tot}}\bigg[1+\frac{\Gamma_{\rm ICM}}{\Gamma_{\rm
      b}}\frac{(\Gamma_{\rm b}-1)}{(\Gamma_{\rm ICM}-1)}\bigg]
\\ \nonumber + \frac{g}{w}\frac{(\rho_{\rm tot}-(1+k)\rho_{\rm
    ICM})}{\rho_{\rm tot}} \\ \nonumber - \frac{(\Gamma_{\rm b}-1)\mu
  m_{\rm p} w g}{\Gamma_{\rm b}k_{\rm B}T_{\rm ICM}} \frac{k \rho_{\rm
    ICM}}{\rho_{\rm tot}} \\ \nonumber - \frac{F}{V_{\rm bub} w
  \rho_{\rm tot}}.
\end{eqnarray}
where $V_{\rm bub} $ is the bubble volume, $k\rho_{\rm ICM}V_{\rm
  bub}$ is the drift mass and defining a total effective density
\begin{equation}
\rho_{\rm tot} \equiv \rho_{\rm bub}+k\rho_{\rm ICM}+ \frac{qM_{\rm
    dis,0}}{V_{\rm bub}},
\end{equation}
with $q M_{\rm dis,0}$ being the wake mass.

The terms in equation (\ref{eq:38}) represent entrainment, buoyancy,
added mass and drag, respectively. A further brief inspection reveals
that the adiabatic relations are recovered in the limit that
entrainment is unimportant ($\alpha \rightarrow 0$). The effective
adiabatic index of the fluid mixture within the bubble is given by
equation (\ref{eq:upgammax}).

In addition to these equations, we must also specify functional forms
for the ICM density and temperature, which can be used to calculate
the background gravitational field. As an example, the ICM density can
be modelled as a $\beta$-profile, e.g. equation (\ref{eq:5}).  For the
temperature profile, we use the Gaussian function used by
\cite{ghizzardi04} to describe the ICM temperature of the Virgo
cluster
\begin{equation}
T_{\rm ICM} = T_1 - T_2 \exp\bigg(-\frac{1}{2}\frac{z^{2}}{z_{\rm
    T}^{2}}\bigg), 
\end{equation}
where $T_1$ is the ICM temperature at large radii, and $T_1 - T_2$ is
the temperature in the centre of the cluster. $z_{\rm T}$ is the scale
height of the temperature profile.

The background gravitational acceleration of the cluster potential is
determined from the temperature and density profiles by assuming the
gas is in hydrostatic equilibrium
\begin{equation}
g = \frac{k_{\rm B}}{\mu m_{\rm p}}\bigg(\frac{{\rm d}T_{\rm
    ICM}}{{\rm d}z} + \frac{T_{\rm ICM}}{\rho_{\rm ICM}}\frac{{\rm
    d}\rho_{\rm ICM}}{{\rm d}z}\bigg)
\end{equation}

Within this framework, we take a fit from \cite{pope06} to the density
profile of the Hydra cluster based on the data given by
\cite{davhyd01}. The values are representative of general cluster
density profiles and are $\rho_0 = 0.7 \times 10^{-25}\, {\rm
  g\,cm^{-3}}$, $\beta=0.72$ and $z_{\rm ICM,0} = 18.6\,$kpc. The
temperature is taken to be a typical cool core cluster in which the
central temperature is a factor of 3 lower than at large radii: $ T_1
= 3 \times 10^{7}\,$K, $T_2 = 2 \times 10^{7}\,$K with $z_{\rm T} =
5.0\,$kpc.

To illustrate the model, we have numerically integrated equations
(\ref{eq:36}), (\ref{eq:37}) and (\ref{eq:38}) using a standard
fourth-order Runge-Kutta method. For the initial conditions, the
bubbles were placed 10 kpc from the cluster centre, with a radius of 5
kpc and a density contrast of $\approx 40$. In both entraining and
non-entraining cases, the added mass coefficient was set to $k=0.5$,
with $q = 0.24$. The initial adiabatic index of the bubble is
$\Gamma_{\rm b,0} = 4/3$, $\Gamma_{\rm ICM} = 5/3$ and $C_{\rm D} =
0.5$. By definition, $\alpha=0$ in the non-entraining case, while for
the entraining bubble we set $\alpha = 0.15$, as expected for mixing
due to Rayleigh-Taylor instabilities.

Figure 3 shows the density contrast of bubbles evolving in the
Hydra-like cluster atmosphere. In these plots, the solid line
represents the non-entraining bubble, while the long and short dashed
lines indicate the behaviour of the entraining bubble with and without
a wake, respectively.

\begin{figure*}
\centering
\includegraphics[width=10cm]{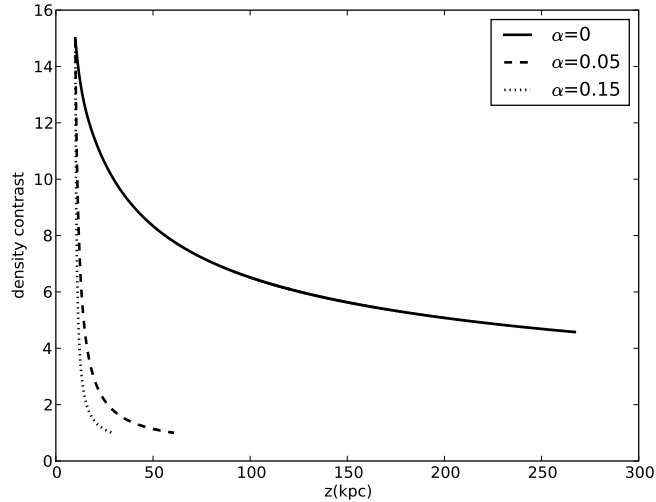}
\caption{Plot showing the density contrast for the entraining bubble
  ($\alpha = 0.15,0.05$) and a non-entraining bubble ($\alpha = 0$).
  Note that the density contrast rapidly becomes small in the
  entraining cases. In fact, the final density contrast is less than
  unity because $\Gamma_{\rm b}$ is always less than $\Gamma_{\rm
    ICM}$, see equation (\ref{eq:37}).}
\label{fig:flow1}
\end{figure*}

As can be seen from figure 3, it is clear that entrainment increases
the rate of expansion of the bubble over and above what would be
expected due to adiabatic expansion. According to equation
(\ref{eq:36}), for large values of $\alpha$ the half-opening angle of
a bubble would tend to
\begin{equation}\label{eq:56} 
\tan(\theta) \equiv \frac{{\rm d}r}{{\rm d}z} \sim \frac{5}{8}\alpha,
\end{equation}
using $\Gamma_{\rm b}=4/3$ and $\Gamma_{\rm ICM} = 5/3$.

\subsection{Entrainment: Observational implications}

As already discussed in the context of the bubble wake, bubble
dynamics and energetics both depend on the density contrast of a
rising bubble.  Entrainment obviously contributes to the lowering of
this contrast.  In addition to this, understanding the impact of
entrainment is of particular interest because \cite{diehl28} have
noted that bubble radii are approximately half their distance from the
cluster centre.  They argue that this correlation is too large and too
constant to be attributable to simple adiabatic
evolution. \cite{diehl28} favour a model where bubbles are inflated by
current-dominated jet \citep[][]{nakamura}.  \cite{brug}, however,
argue that this behaviour may well be explicable by a combination of
projection effects and modifying the standard adiabatic model to
include the influence of entrainment.

According to observations, the typical opening angle of a bubble is
roughly $\tan(\theta) \sim 0.5$, \citep[e.g.][]{diehl28}. If
entrainment is the physical process which leads to this, equation
(\ref{eq:56}) implies that $\alpha\sim 0.8$.  This, in turn, implies
an entrainment length-scale of $L \sim r_0/2$.  In other words, the
bubble could only rise a distance equivalent to the initial bubble
radius before it became indistinguishable from its surroundings.  It
is difficult to rule out this possibility entirely since we cannot
know the bubble radii at earlier times.  However, bubbles clearly
persist out to large distances - the cavity statistics provided by
\cite{birzan} indicate that there are bubbles located more than 20 kpc
from the cluster centre in Perseus, A4059, A133, A2199, RBS 797,
A2597, Hydra A, Cygnus A.

Moreover, \cite{sand} argue, based on an analysis of the bubbles in
Perseus, that the volume fraction of thermal material in bubbles
cannot exceed 50\%.  The conversion of the volume fraction into a mass
fraction depends requires making certain assumptions about, for
example, the temperature of the entrained material and whether it is
heated significantly as it enters the bubble.  Nonetheless, it is
possible to derive some general expectations: if the entrained
material is at the ambient temperature then the density contrast of
the rising bubbles must be high and the entrainment process must be
very inefficient (i.e.~$\alpha$ must be small), otherwise the bubbles
would not appear as depressions in the X-ray surface brightness
maps\citep[e.g.][]{mcnuls}.  If the entrained gas is heated above 15
keV, it would be difficult to detect using existing X-ray telescopes
and therefore, the appearance of AGN bubbles as depressions in the
X-ray surface brightness maps offers no constraint.  On the other
hand, if the bubble and the ICM are in pressure equilibrium, then the
fraction of the bubble mass attributed to entrainment depends roughly
on the ratio of the ambient temperature to that of the newly heated
material in the bubble \citep[e.g.][]{mc08}.  If the entrained
material is heated significantly, the entrained mass can again not be
very large.  For example, if the entrained material is heated to be,
say, five times the temperature of the ambient gas, the bubble will
contain at most one fifth of the mass originally displaced by the
inflation process.

Consequently, the upper limit of 50\% for the volume fraction of the
thermal content strongly suggests that the bubbles cannot be
entraining rapidly and the large opening angles must have a different
explanation.  This can be illustrated using the similar reasoning used
to derive equation (\ref{eq:upgamma}).  Specifically, the pressure
fraction, $f_{\rm i+1}$, contributed by entrained material after the
$i$th timestep of the numerical integration using
\begin{equation}
\frac{1}{\Gamma_{\rm i+1}-1} = \frac{1-f_{\rm i+1}}{\Gamma_{\rm
    b,0}-1} + \frac{f_{\rm i+1}}{\Gamma_{\rm ICM}-1},
\end{equation}
where $\Gamma_{\rm b,0}$ is the initial adiabatic index of the bubble
before it entrained any material from its surroundings and $f_{\rm
  i+1} = {\rm thermal\, pressure/total\, pressure}$. Note that this is
the volume-averaged pressure fraction. Therefore, since the pressures
of each component must have been equal, and given that material is
entrained at constant pressure, the volume-averaged pressure fraction
is equivalent to the volume fraction of thermal material.

\begin{figure*}
\centering
\includegraphics[width=10cm]{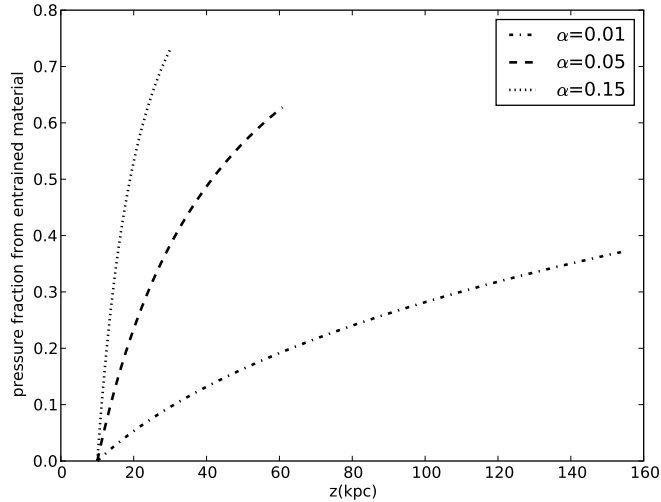}
\caption{Plot showing the fraction of internal bubble pressure
  contributed by entrained material for different entrainment rates:
  $\alpha=0.15,0.05,0.01$. By definition, the pressure fraction for
  the non-entraining bubble is always zero.}
\label{fig:flow3}
\end{figure*}

Figure 4 clearly shows that, for $\alpha = 0.15$, the thermal pressure
fraction grows rapidly to greater than 0.6 before the bubble has risen
10 kpc from its original location. However, for $\alpha = 0.05$, the
pressure fraction is 0.5 at roughly 40 kpc from the cluster centre and
rises much more slowly. For $\alpha = 0.01$, the pressure fraction
remains $<$50\% even beyond 150 kpc. Consequently, a thermal volume
fraction of $\sim 50\%$ suggests either that the thermal content was
injected (or entrained) during the initial inflation, or that the
subsequent entrainment rate is no larger than $\alpha \approx
0.05$. This reduced rate probably indicates that magnetic fields at
the bubble/ICM interface suppress the growth of fluid instabilities,
while still permitting a small amount of thermal material to enter the
bubble due to cross-field diffusion and magnetic reconnection events
\citep[e.g.][]{pope2010}.  In any case, the large opening angles of
the bubbles cannot be due to entrainment.  Alternative possibilities
include:\hfill\break (1) The bubbles are hydrodynamic in origin and
the apparently large opening angle is partially due to projection
effects, as suggested by \cite{brug};\hfill\break (2) The bubbles are
{\em not} hydrodynamic in nature and are formed by a different
inflation process, e.g. current-dominated jets, as suggested by
\cite{diehl28}.

\section{Discussion}

Having examined the three main mass transport processes associated
with a buoyant bubble individually, we now consider the three
processes jointly and ask:\hfill\break (1) how much material in total
can actually be transported per bubble?\hfill\break (2) how frequently
must bubbles be inflated in order for the mass outflow to balance the
cooling inflow?

Regarding the first question, we have already established that the
bubbles cannot be carrying very much entrained material.  The drift
can carry upward an amount of ambient material comparable to 50\% of
the mass initially displaced by the bubble, or approximately $\sim
10^{8}\,{\rm M_\odot}$.  However, most of this material will not be
displaced upwards by more than a bubble diameter and will likely fall
back towards the cluster centre.  In addition, for clusters with a
flat entropy core, the overall fluid displacement might be negligible.
This then means that the drift, though likely responsible for
observable features like the cool optical filaments, is not an
important mechanism for affecting mass outflow.

This leaves the wake.  As noted previously, fluidised bed experiments
\citep[e.g.][]{yang} show that wake typically contains a approximately
a quarter of the mass initially displaced by the bubble.  Assuming
that laboratory value of $q$ can be applied to the ICM, AGN-blown
bubble wakes could easily transport $q M_{\rm dis,0} \sim 10^{8}\,{\rm
  M_\odot}$, see equation (\ref{eq:disp}).

Moreover, the mass in the wake can be carried out to large distances.
As indicated by equation (\ref{eq:6a}), the maximum bubble height is a
strong function of $q$. For $q \sim 0.24$ (and $\eta_{0} = 100$),
depending on the temperature profile, the bubble can rise to at least
5$z_{\rm 0}$, where $z_{0}$ is the height at which the bubble was
inflated.  For $z_{0}\sim 10$ kpc, the bubble will rise to 50 kpc.
 
Let us consider an AGN that inflates a sequence of bubbles.  We can
write the time-averaged mass outflow rate from the cluster centre due
to a train of bubbles as
\begin{equation}\label{eq:beg}
\dot{M}_{\rm out}\approx qM_{\rm dis,0}/\tau,
\end{equation}
where $\tau$ is the average time between bubble inflation.  Since this
outflow counteracts the inflow due to cooling, cumulatively bubble
mass transport may allow for the relaxing of the stringent (but
commonly adopted) requirement that AGN heating must balance radiative
cooling.  In fact, there is some debate as to whether AGN heating in
fact balances cooling, especially in the most luminous clusters.
Estimates of the time-averaged heating rate by \cite{best08}, for
example, indicates that AGN heating {\em does not} balance cooling in
the most luminous clusters.\footnote{We note that \cite{dunn08} claim
  a close balance between heating and cooling but in their comparison,
  they refer to the heating rate due to an individual bubble.
  Averaging over the time between successive bubbles will lower the
  power estimate, leading to a discrepancy between heating and cooling
  in agreement with \cite{best08}.}  As we show below, mass outflow
can play an important role in limiting the mass build-up at the
cluster centre in such instances.
 
To illustrate this, let us consider the following: in the absence of
heating, the mass inflow rate due to cooling is related to the
radiative losses of the ICM according to
\begin{equation}
\dot{M}_{\rm cool,0} =  \frac{(\Gamma_{\rm ICM}-1)}{\Gamma_{\rm ICM}}
\frac{\mu m_{\rm p}}{k_{\rm B}T} L_{\rm X},
\end{equation}
where $T$ is the average temperature of the ICM.  Studies of the X-ray
spectra of the ICM in cluster cores indicate that the actual rate at
which the gas is cooling is $\dot{M}_{\rm cool}\simlt 0.2 \dot{M}_{\rm
  cool,0}$ \cite[e.g.][]{peterson01,kaas01,peterson03}. If the mass
inflow rate is balanced by the mass transport rate by the bubbles:
$\dot{M}_{\rm cool}\approx \dot{M}_{\rm out}$.

The actual gas cooling rate, $\dot{M}_{\rm cool}$, is the result of a
difference between cooling and heating:
\begin{equation}\label{eq:end}
\dot{M}_{\rm cool} = \frac{(\Gamma_{\rm ICM}-1)}{\Gamma_{\rm ICM}}
\frac{\mu m_{\rm p}}{k_{\rm B}T} \left(L_{\rm X}-\dot{E}_{\rm
  heat}\right),
\end{equation}
where $\dot{E}_{\rm heat} = \alpha_{\rm heat} E_{\rm bub,0}/\tau$.
Here, $E_{\rm bub,0}$ is the enthalpy of an AGN-inflated bubble (see
equation \ref{eq:1}) and $\alpha_{\rm heat}$ allows for the
possibility that the bubble enthalpy may under-represent the total
energy injected by the AGN \citep[e.g.][]{nusser,binney07}, especially
if the bubbles are inflated supersonically, which would engender
shocks that enhance the heating.

Combining equations (\ref{eq:beg})---(\ref{eq:end}) and taking
$\Gamma_{\rm ICM} = 5/3$, $\Gamma_{\rm b} = 4/3$ and $T/T_0 \sim 3$,
we find that the build-up of excessive amounts of cold gas in the
central galaxy can be prevented if
\begin{equation}
\dot{E}_{\rm heat} \approx \frac{\alpha_{\rm heat} }{\left[
    2q+\alpha_{\rm heat}\right]}L_{\rm X}
\end{equation}
and
\begin{equation}
\dot{M}_{\rm out} \approx \frac{2q }{\left[2q+\alpha_{\rm
      heat}\right]}\dot{M}_{\rm cool,0}.
\end{equation}
If the AGN does no heating (i.e.~$\alpha_{\rm heat}=0$), mass
transport by the wake can, by itself, prevent rate of cold gas in the
central cluster galaxy as long as the bubble recurrence timescale is
\begin{equation}
\tau \approx 1 \times 10^{7} \bigg(\frac{E_{\rm bub,0}}{10^{59}\,{\rm
    erg}}\bigg)\bigg(\frac{L_{\rm X}}{10^{44}\,{\rm
    erg\,s^{-1}}}\bigg)^{-1}\,{\rm yrs},
\end{equation}
where $10^{59}\,{\rm erg}$ is the typical enthalpy of bubbles in a
cluster with luminosity $L_{\rm X} \sim 10^{44}\,{\rm erg s^{-1}}$
\citep[e.g.][]{birzan,dunn05} and we have taken $q\approx 0.24$.  Of
course, the corresponding $\dot{M}_{\rm cool}$ will exceed the limits
derived from the X-ray spectra.

If, in addition to mass transport, the bubbles also heat the ICM such
that the amount of energy deposited in the ICM is twice the enthalpy
of the bubble (i.e.~$\alpha_{\rm heat}=2$), the rate of cold gas
build-up can be prevented if the bubbles are inflated on a timescale
\begin{equation}\label{eq:tcritx}
\tau \approx 5 \times 10^{7} \bigg(\frac{E_{\rm bub,0}}{10^{59}\,{\rm
    erg}}\bigg)\bigg(\frac{L_{\rm X}}{10^{44}\,{\rm
    erg\,s^{-1}}}\bigg)^{-1}\,{\rm yrs}.
\end{equation}
It is worth noting that in this particular case, the time-averaged
heating rate does not balance the cooling rate, $\dot{E}_{\rm
  heat}\approx 0.8 L_{\rm X}$, and the mass cooling rate is the
maximum allowed, $\dot{M}_{\rm cool} \approx 0.2\dot{M}_{\rm cool,0}$,
but mass build-up in the central cluster galaxy is successfully
avoided by mass transport due to the bubbles.

The values of $\tau$ obtained from equation (\ref{eq:tcritx}) are reasonably consistent with indirect observational estimates, though there are large uncertainties. For example, \cite{best08} found radio AGN duty cycles of $\sim 30\%$ in Brightest Cluster Galaxies (BCGs), which corresponds to an  typical `observed' time between outbursts of $\tau_{\rm obs} \sim 3 t_{\rm on}$, where $t_{\rm on}$ is the duration of an AGN outburst. Therefore, since $t_{\rm on}$ seems to vary between $10^{7}-10^{8}\,{\rm yrs}$ \citep[e.g.][]{birzan}, it seems possible that AGN-blown bubbles can significantly reduce the rate at which cold gas collects in the galaxy.

Finally, we note that throughout this work we have only considered
bubbles that remain intact. Bubbles which deform into vortex rings
behave somewhat differently. In addition, the fluid circulation
associated with the vortex ring exerts a force that can lift up
additional material behind the bubble, \citep[see][]{pav}. However,
with the possible exception of M87 \citep[][]{m87rad} structures
resembling vortex rings have not been observed, so this additional
mechanism of mass transport might be unimportant except in rare cases.

\section{Summary}

The aim of this article is to provide a better understanding of the
mass transport processes that are attributable to AGN-blown bubbles in
clusters of galaxies.  Recently observed filaments of cool material
behind AGN-blown bubbles \citep[e.g.][]{consel,crawf05,hatch} provide
a strong indication that bubble-induced mass transport is operating at
least within 10s of kiloparsecs of the cluster centre.  The filaments
likely represent only a fraction of the actual mass transported by the
bubbles and some of this mass can be carried out to $\sim 50-100$
kiloparsecs.

Mass transport by AGN-inflated bubbles not only impacts bubble
dynamics --- including the rate at which the bubble expands as it
rises --- and the efficiency with which the bubble will heat its
surroundings, it also removes the amount of cool gas available for
star formation in the central galaxy, alleviating the stringent
requirement that AGN heating balance radiative cooling. The system can
tolerate a lower heating rate, implying that the systems identified by
\cite{best08}, in which the average heating rates fall short of the
cooling rates, may still be balanced.  For all these reasons, a better
understanding of mass transport by the bubbles is important.  In this
paper, we have focused on three main mass transport mechanisms: drift,
wake transport and entrainment.
\begin{itemize}
\item{\bf Drift:} drift arises when an ascending bubble imparts an
  impulse on the ambient ICM ahead of it, causing a net upward
  displacement of the ambient material behind the bubble.  In a
  gravitationally stratified environment, the material in the drift
  does not keep up with the bubble.  It slows down and may even fall
  back to its original position.
\item{\bf Wake transport:} a rising, deformable bubble also forms a
  cavity (wake) at the rear of the bubble.  The wake is filled with
  some of the ambient medium displaced when the bubble first starts to
  rise, and this material will be transported outward as the bubble
  rises.
\item{\bf Entrainment:} during formation and possibly, during the rise
  phase, some of the ambient ICM may be incorporated into the bubble
  body.  This process is referred to as entrainment.  The displacement
  of the bubble to larger radii in the cluster results in the
  transport of this entrained material.
\end{itemize}

Our most important findings regarding the above three processes can be
summarized as follows:
\begin{enumerate}
\item The upward displacement of the cool gas at the cluster centre in
  the form of drift offers a compelling explanation for the recently
  discovered cool filaments observed to be stretching between
  AGN-inflated bubbles and the cluster centre.  The kinematic
  character of the filaments agrees well with the expected kinematic
  signature of drift.  Our estimate of a drift mass of $\sim
  10^{8}\,{\rm M_\odot}$ also agrees well with the estimate of mass in
  the cool filaments.  We generically expect such filaments to form in
  clusters that either have entropy profiles that increase with radius
  (as is common in classical cooling core clusters) or had such an
  entropy profile at the time when the bubble started its ascent.
\item The wake is the most promising mechanism for transporting
  material out of the cluster centre and out to distances of $\sim
  100$ kpc, depending on where the bubble is inflated and the ICM's
  radial density profile.  (In the absence of the wake, the
  equilibrium radius of the very same bubble greatly exceeds the
  cluster's virial radius.)  Our estimates suggest that typically a
  bubble will lift $\sim 10^{8}\,{\rm M_\odot}$ in its wake, which
  significantly weighs down the bubble.  A hot air balloon with a
  weighted basket is an apt metaphor.  This additional mass not only
  limits the maximum radius to which a bubble can rise but also
  significantly reduces the energy that can be extracted from the
  bubble.  In a classical cool core cluster, a weighted bubble will
  deposit into the ICM only $\sim 18\%$ of the energy released
  associated with an unweighted bubble.
\item \cite{brug} recently suggested that entrainment and projection
  effects may provide an explanation for the large apparent opening
  angles subtended by AGN-inflated bubbles. The presence of entrained
  material {\em does} increase the bubble's expansion rate.  However,
  the entrainment rate required to account for the observations would
  also rapidly reduce the density contrast between the bubble and its
  surroundings. In a typical cluster, a bubble would only need to rise
  a distance equivalent to the initial bubble radius before it would
  become indistinguishable from its surroundings.  Since \cite{birzan}
  find numerous bubbles located more than 20 kpc from the cluster
  centres, we conclude that the bubble does not entrain material
  efficiently.  There are two related implications: firstly, the large
  opening angle must be due to some other process, such as
  current-dominated AGN outflow model \citep[e.g.][]{nakamura} as well
  as possible projection effects, as suggested by \cite{brug}.
  Secondly, some phenomenon prevents the formation of normal
  hydrodynamic instabilities and, therefore, suppresses entrainment.
  As a possible resolution, we note that bubble whose surface is
  draped by magnetic fields would behave as required.  Whether such a
  bubble would be consistent with other constraints is a subject for
  further explorations.
\item With the introduction of mass transport of gas out of the
  cluster centre by the bubble, we find the clusters of galaxies can
  tolerate an average heating rate that is less than the cooling rate,
  which according to \cite{best08} is a common occurrence in luminous
  clusters, and prevent gas build-up in the central galaxy.  The
  outward mass transport rate is largely determined by the wake mass
  and by the average time between bubble inflation.  Typically, the
  bubble recurrence timescale is $\tau \sim 10^7$---$10^8$ yrs, which is reasonably consistent with observational estimates.
\end{enumerate}

\section{Acknowledgements}

ECDP would like to thank the Department of Foreign Affairs and
International Trade for funding through a Government of Canada
Post-Doctoral Research Fellowship and CITA for funding through a
National Fellowship. He would also like to acknowledge informative
discussions with Jim Hinton and Ian Eames. GP thanks the STFC for
financial support. AB acknowledges research support from NSERC through
the Discovery Program and also expresses his gratitude to J. Criswick
for his generous support. AD thanks CITA for financial support through
a National Fellowship and an NSERC grant to Don VandenBerg is also
acknowledged. The authors wish to thank the anonymous referee for
insightful comments that improved this work.

\appendix

\vfill

\section{List of parameters and their definitions}\label{app:param}

\begin{center}
\begin{tabular}{p{1in}p{2in}}
  \hline
  Parameter & Definition\\
  \hline
  $E_{\rm bub,0}$ & bubble enthalpy immediately after inflation \\
  $P$ & pressure (bubble and ICM are in pressure equilibrium)\\
  $P_{0}$ & pressure where bubble was inflated\\
  $V_{\rm bub}$ & bubble volume\\
  $V_{\rm bub,0}$ & initial bubble volume\\
  $\rho_{\rm bub}$ & bubble density\\
  $\rho_{\rm bub,0}$ & initial bubble density\\
  $T_{\rm bub}$ & bubble temperature\\
  $\Gamma_{\rm b}$ & bubble adiabatic index\\
  $\Gamma_{\rm b,0}$ & initial bubble adiabatic index\\
  $\rho_{\rm ICM}$ & ICM density\\
  $\rho_{\rm ICM,0}$      & central ICM density\\
  $\rho_0$ & ICM density where the bubble was inflated\\
  $T_{\rm ICM}$ & ICM temperature\\
  $T_0$       & ICM temperature where bubble was inflated\\
  $T_{1},T_{2},z_{\rm T}$     &  parameters for fit to ICM temperature profile\\
  $\Gamma_{\rm ICM}$ & ICM adiabatic index\\
  $\eta_0$ & defined as $\rho_0/\rho_{\rm bub,0}$\\
  $\eta_{\rm eff}$ & effective density contrast for bubble and wake system\\
  $\eta_{\rm eff,0}$ & effective density contrast for bubble and wake system at the beginning of the rise phase\\
  $\beta$ & exponent for $\beta$-law ICM density\\
  $z$ & radial coordinate of cluster\\
  $z_{\rm ICM,0}$ & scale height of $\beta$-law ICM density\\
  $z_0$ & height at which the bubble is inflated\\
  $z_{\rm eq}$ & height at which effective density of the bubble/wake systems equals the ICM density\\
  $P(z_{\rm eq})$ & pressure at $z_{\rm eq}$ \\
  $T_{\rm ICM}(z_{\rm eq})$ & ICM temperature at $z_{\rm eq}$ \\  
  $V_{\rm drift}$ & drift volume\\
  $V_{\rm wake}$ & wake volume\\
  $M_{\rm bub}$ & bubble mass\\
  $M_{\rm dis}$ & ICM mass displaced by bubble\\
  $M_{\rm dis,0}$ & initial ICM mass displaced by bubble\\
  $M_{\rm wake}$ & mass of bubble wake\\
  $q$ & fraction of initial displaced mass in the wake\\
  $k $ & ratio of drift and bubble volumes\\
  $g$ & gravitational acceleration\\
  $w$ & bubble velocity\\
  $w_{\rm terminal}$ & terminal bubble velocity\\
  $r$ & bubble radius\\
  $r_0$ & initial bubble radius\\
  $a_{\rm bub}$ & bubble surface area\\
  $\epsilon$ & turbulent energy dissipation rate per unit mass\\
  $N$ & Brunt-V\"ais\"al\"a frequency\\
  \hline
\end{tabular}
\end{center}

\begin{center}
\begin{tabular}{p{1in}p{2in}}
  \hline
  Parameter (cont...) & Definition\\
  \hline
  $\sigma$ & entropy index of ICM\\
  $C_{\rm D}$ & drag coefficient for bubble\\
  $L_{\rm O}$ & Ozmidov length\\
  $L_{\rm buoy}$ & buoyancy length\\
  $L$ & entrainment length\\
  $c_{\rm turb}$ & numerical turbulence constant\\
  $c_{\rm term}$ & drag coefficient-dependent  \\
  $Re$ & Reynolds number\\
  $n$ & number of computational cells\\
  $\alpha$ & entrainment coefficient\\
  $\Gamma_{\rm i}$ & current bubble adiabatic index\\ 
  $f_{\rm i}$ & fraction of bubble pressure contributed by entrained material\\
  $\Delta V$ & volume increment of entrained material\\
  $\Delta M$ & mass increment of entrained material\\
  $\Delta V_{\rm i}$ & volume entrained during $i$th timestep\\
  $V_{\rm i}$ & bubble volume in $i$th timestep\\
  $\Gamma$ & effective adiabatic index\\
  $c_{\rm p,bub}$ & specific heat at constant pressure of bubble\\
  $c_{\rm p,ICM}$ & specific heat at constant pressure of ICM\\
  $\theta$ & opening angle of bubble\\
  $L_{\rm X}$ & X-ray luminosity of ICM\\
  $T$ & characteristic temperature of ICM\\
  $\dot{M}_{\rm cool,0}$ & classical mass flow rate associated with $L_{\rm X}$ and $T$\\
  $\dot{M}_{\rm cool}$ & the actual cooling flow\\
  $\alpha_{\rm heat}$ & multiple of bubble enthalpy that represents total energy injected by AGN\\
  $\tau$ & average time between bubbles being inflated\\
  $\tau_{\rm obs}$ & observationally-determined time between bubbles being inflated\\
  $t_{\rm on}$ & duration of an AGN outburst\\
  $k_{\rm B}$ & Boltzman constant\\
  $\mu m_{\rm p}$ & mean mass per particle\\
  $A$ & cross-sectional area of bubble\\
  \hline
\end{tabular}
\end{center}



\bibliography{database} \bibliographystyle{mn2e}

\label{lastpage}

\end{document}